\documentstyle[prl,aps,psfig]{revtex}
\begin{document}
\draft
%
\twocolumn[
\title{
Spin Dynamics in an Ordered Stripe Phase}
\author{J. M. Tranquada,$^1$ P. Wochner,$^1$ and D. J.
Buttrey$^2$}
\address{$^1$Physics Department, Brookhaven National Laboratory, Upton, 
New York 11973}
\address{$^2$Department of Chemical Engineering, University of Delaware,
Newark, Delaware 19716}
\date{December 2, 1996}
\maketitle
\widetext
\advance\leftskip by 57pt
\advance\rightskip by 57pt
\begin{abstract}
Inelastic neutron scattering has been used to measure the low-energy spin
excitations in the ordered charge-stripe phase of La$_2$NiO$_{4+\delta}$ with
$\delta=0.133$.  Spin-wave-like excitations disperse away from the
incommensurate magnetic superlattice points with a velocity $\sim60$\%\ of that
in the $\delta=0$ compound.  Incommensurate inelastic peaks remain
well-resolved up to at least twice the magnetic ordering temperature. 
Paramagnetic scattering from a $\delta=0.105$ sample, which has a
N\'eel-ordered ground state, shows anomalies suggestive of incipient stripe
correlations.  Similarities between these results and measurements on
superconducting cuprates are discussed.		
\end{abstract}
]
\narrowtext

Hole-doped La$_2$NiO$_4$ is a strongly-correlated electron system that exhibits
an exotic form of charge order.  Although many questions concerning this order
remain to be answered, the basic nature of the charge and associated spin order
has now been fairly well established by scattering
studies \cite{hayd92,chen93,yama94,tran94a,woch97,sach95}. 
At sufficiently high levels of doping ($\gtrsim 0.15$ holes/Ni), the added holes,
which enter the NiO$_2$ planes, order in periodically spaced stripes.  The Ni
spins in the intervening regions order antiferromagnetically, with an antiphase
relationship between neighboring domains.  To make further progress toward
understanding  the spatially-inhomogeneous charge and spin correlations in this
system, it is necessary to study the spin and charge dynamics.  Some information
on charge dynamics has been provided by infrared reflectivity
studies \cite{ido91}.  Here we present an inelastic
neutron scattering study that compares the spin dynamics in 3 distinct phases of
La$_2$NiO$_{4+\delta}$, one with a stripe-ordered ground state ($\delta=0.133$)
and two with N\'eel-ordered ground states ($\delta=0$ and 0.105).

Static stripe order has also been observed in the cuprate system
La$_{1.6-x}$Nd$_{0.4}$Sr$_x$CuO$_4$ \cite{tran95a}, where it has been 
shown to coexist with superconductivity \cite{tran97a}.  The {\bf Q} dependence
of the magnetic scattering is essentially identical with that of the dynamical
spin correlations found in La$_{2-x}$Sr$_x$CuO$_4$ \cite{cheo91}.  Given the
empirical similarities between stripe-ordered phases in the nickelates and
cuprates, it is essential to explore similarities and differences of the spin
dynamics in these two classes.  Such comparisons are made throughout
the paper.

Some characterization of spin excitations associated with stripe order was
reported by Hayden {\it et al.}\cite{hayd92} in their pioneering study of
incommensurate magnetic correlations in a sample of Sr-doped La$_2$NiO$_4$;
however, the short spin-spin correlation length associated with the dilute and
randomly distributed dopants limited the information that could be obtained. 
Oxygen-doped samples have the advantage that for $\delta\gtrsim0.11$, the
interstitials and the charge stripes both order
3-dimensionally \cite{yama94,tran94a,woch97}.  Previous
neutron-scattering studies of spin dynamics in La$_2$NiO$_{4+\delta}$
\cite{aepp88,yama91,naka93,naka95} have focussed on samples with $\delta$
in the range $0\le\delta\lesssim0.11$, throughout which the magnetic order
remains commensurate, albeit with a strong variation in ordering temperature,
$T_N$ \cite{hoso92,tran94b}.  The reduction in
$T_N$ with increasing $\delta$ is accompanied by a dramatic decrease in the
spin-wave velocity, the change being a factor of $\sim3$ between $\delta=0$ and
0.077 \cite{aepp88,yama91,naka93}.  Above $T_N$, the instantaneous magnetic
correlation length is found to decay with temperature much more rapidly at
higher $\delta$ \cite{naka95}.  

A simple extrapolation of the lower $\delta$ results might lead one to expect
that an increasing hole density would result in a severe reduction in the
effective magnetic interaction strength.  To the contrary, we will demonstrate
that at higher
$\delta$, where the holes order in charge stripes, the spin waves, dispersing
away from incommensurate wave vectors, actually harden.  The incommensurability
of the low-energy spin fluctuations remains well resolved up to at least twice
the magnetic ordering transition, $T_m$.  Furthermore, we will show that,
despite commensurate order below $\sim50$~K, the paramagnetic scattering from a
$\delta=0.105$ crystal is inconsistent with the behavior expected for a quasi-2D
Heisenberg system such as La$_2$NiO$_4$ \cite{naka95b}.  The development of a
temperature- and energy-independent lineshape for $T\gtrsim120$~K is
suggestive of incipient charge-stripe correlations.

The crystals of La$_2$NiO$_{4+\delta}$ with $\delta=0$, 0.105, and 0.133
discussed here have been characterized elsewhere \cite{woch97,tran94b}.  The
interstitials in the $\delta=0.105$ sample (and throughout the range
$0.05\lesssim\delta\lesssim0.11$) exhibit a 1-dimensional staging order
\cite{tran94b}, whereas those in the $\delta=0.133$ sample order 3-dimensionally
\cite{woch97}.  The change in the nature of the interstitial order is
correlated with the change in character of the spin and charge order.

Inelastic neutron scattering measurements were performed on triple-axis
spectrometers at Brookhaven National Laboratory's High Flux Beam Reactor. 
Neutrons were monochromatized and analyzed with pyrolytic graphite (PG)
crystals set for the (002) reflection.  The final neutron energy was fixed at
14.7~meV, and a PG filter was used to eliminate higher order neutrons from the
scattered beam.  Coarse horizontal collimations were selected
($40'$-$40'$-$80'$-$80'$ for most of the measurements).  Each sample was cooled
with a Displex closed-cycle He refrigerator.

To describe the results we will make use of a unit cell of
size $\sqrt{2}{\bf a}_t\times\sqrt{2}{\bf a}_t\times{\bf c}_t$ relative to the
simple tetragonal one.  The antiferromagnetic wave vector for a single NiO$_2$
layer is then $(1,0,0)$, or equivalently
$(0,1,0)$, where reciprocal lattice vectors {\bf Q} are specified in units of
$(\frac{2\pi}a,\frac{2\pi}a,\frac{2\pi}c)$.  The spin fluctuations in the
stripe-phase sample were measured in the $(h,k,0)$ zone near $(0,1,0)$.  To
simplify comparison, the dispersion results for the N\'eel-ordered samples will
also be presented in the $(h,k,0)$ zone, although they were actually measured
in a different zone.

To optimize the resolution and cross section, spin-wave measurements of the
$\delta=0$ and $\delta=0.105$ samples were performed in the $(h,0,l)$ zone,
either by scanning $h$ at fixed energy transfer or by
scanning the energy transfer at fixed $h$, with $l$ selected to minimize the
effective resolution width.  The results obtained at low temperature
($\sim10$~K) are indicated by the circles ($\delta=0$) and triangles
($\delta=0.105$) in Fig.~1.  The lines through the data are spin-wave dispersion
curves with the spin-wave velocities, $\hbar c$, as shown in Table I.  The
velocity for
$\delta=0$ is taken from Ref.~\cite{yama91}, whereas that for the 0.105 sample
is a fit to the data.  The two curves for each $\delta$ correspond to in-plane
and out-of-plane spin-wave modes, which have different anisotropy gaps as
discussed in Ref.~\cite{naka93}.  On increasing
$\delta$ from 0 to 0.105, we find a decrease of the spin-wave velocity by a
factor of 5, consistent with, though more extreme than, the earlier results
\cite{aepp88}.

This trend is abruptly reversed on entering the stripe-ordered phase found at
$\delta=0.133$.  The measurements on this sample were performed in the
$(h,k,0)$ zone, with constant-energy scans through the incommensurate peak at
$(\epsilon,1,0)$.  Scans were
taken both along $(h,1,0)$, in the direction of the modulation, and along
$(\epsilon,k,0)$, parallel to the stripes.  These scan directions are indicated
in the left inset of Fig.~1 as $A$ and $B$, respectively.  The data
collected at 80~K are shown in Fig.~2, with the intensity multiplied by the
energy transfer, $\hbar\omega$, to compensate for the spin-wave-like fall off of
intensity with energy.  If the excitations truly are like spin waves, then we
would expect to see two peaks dispersing away from the incommensurate wave vector
as $\hbar\omega$ increases.  The scans along $B$ in Fig.~2(b) appear consistent
with such a model, and the two peaks are almost resolved at
$\hbar\omega=8$~meV.  The peak separations obtained by fitting
symmetric pairs of Gaussian peaks to the data are shown in the right inset of
Fig.~1, and the resulting spin-wave velocity is listed in Table I.  In
contrast, the scans along $A$ in Fig.~2(a) show a single broad peak at each
energy, with the width increasing linearly with energy. (Resolution does not
limit the width except possibly at 2~meV.) The results of Gaussian fits are shown
in Fig.~1.

Looking at Table I, we see that the spin-wave velocity for excitations
propagating parallel to the stripes is $\sim60$\%\ of that in the $\delta=0$
sample, and 3 times greater than the velocity for $\delta=0.105$.  The decrease
relative to $\delta=0$ reflects the weakened exchange between spins separated
by a charge stripe.  In a simple Heisenberg antiferromagnet, $\hbar c$ is
proportional to the product of the superexchange energy $J$ and the number of
nearest neighbors.  A large part of the decrease in $\hbar c$ can be attributed
to the effective reduction in the number of nearest neighbors.  In contrast,
excitations perpendicular to the stripes appear to have a greater damping,
perhaps due to a more direct coupling to the dynamical spin degrees of freedom
associated with the charge stripes.

We have also measured spin excitations in the charge-ordered, paramagnetic
phase of the $\delta=0.133$ sample at $T>T_m=110.5$~K.  Figure 3 shows scans
along the modulation direction at $\hbar\omega=4$~meV for three different
temperatures.  Although the peak widths increase with temperature, the
incommensurate peaks remain well resolved even at $T=2T_m$. The charge order
is finite but extremely weak at this point \cite{woch97}. 

How do the above results relate to La$_{2-x}$Sr$_x$CuO$_4$?  In the cuprate it
has been shown that the scattering from low-energy spin excitations peak at
incommensurate positions, with the peak widths increasing with both energy and
temperature \cite{cheo91}.  An explanation of how the
incommensurability can be understood in terms of stripe correlations has been
given elsewhere \cite{tran95a}.  The present results demonstrate that the spin
excitations in a stripe-ordered nickelate phase exhibit an energy dispersion and
thermal broadening that is qualitatively similar to that found in the cuprate
system.  Such an interpretation is also consistent with the recent finding that
the high-energy spin excitations in La$_{1.86}$Sr$_{0.14}$CuO$_4$ have a
spin-wave-like character \cite{hayd96}. Of course, there is no static order of
stripes in the cuprates with optimal superconductivity---the dynamic nature of
the stripes is presumably related to the more quantum mechanical nature of the
cuprates.  Nevertheless, it has been shown elsewhere \cite{tran97a} that
superconductivity can coexist with stripe order.

Now let us return to the $\delta=0.105$ sample.  Remember that the Ni moments
order commensurately below $\sim50$~K, and that the spin waves found at low
temperature are strongly renormalized by the doped holes.  If we look at
$T>T_N$, the spin waves are overdamped, and a constant-energy scan through
${\bf Q}_{\rm AF}$ yields a single, broad peak.  The peak widths as a function
of temperature for 3 different energies are indicated on the right-hand side of
Fig.~4.  Initially, the peak widths increase with energy, as well as
temperature, but for $T\gtrsim120$~K the widths become independent of energy
and temperature.  Furthermore, as indicated on the left side of Fig.~4, the peak
shape is rather flat topped.

The energy-independence suggests that the peak shape is determined predominantly
by a modulation of the dynamical spin correlations in real space.  In fact, we
can describe the peak shape quite well with a simple model of antiphase AF
domains with correlations between domains that decay exponentially with
distance.  If we consider domains of width $2a$ (4 Ni spins wide along [010])
as observed in a Sr-doped sample with comparable hole density \cite{sach95},
then the scattered intensity along $(0,k,0)$ should have the form
\begin{equation}
  I = AF^2 {1-p^2\over1+p^2+2p\cos(4\pi k)},
\end{equation}
where $A$ is a scale factor, $p=e^{-2a/\xi}$ with $\xi$ being the
correlation length, and
\begin{equation}
  F = 2\left[\sin\left(\case12\pi k\right)-\sin\left(\case32\pi k\right)\right].
\end{equation}
We get an excellent fit to the data, shown by the curves in Fig.~4, if we
set $p=0.1$, corresponding to $\xi=0.87a=4.7$~\AA.  The weakly correlated
antiphase magnetic domains imply charge segregation.  The disappearance of
these incipient stripe correlations at low temperature contrasts with the case
of La$_{1.8}$Sr$_{0.2}$NiO$_4$ \cite{sach95}. 

These results have implications for two distinct cuprate systems.  First of
all, similar line shapes have been observed in superconducting
YBa$_2$Cu$_3$O$_{6.6}$ \cite{ster94}.  There, the net width of the
flat-topped peaks is smaller than in the nickelate case, implying wider
antiferromagnetic domains.  Nevertheless, the unusual shape suggests the
presence of weakly correlated antiphase domains, as might be induced by
fluctuating charge stripes.  This analogy is
important, because it suggests a uniform interpretation of the spin dynamics in
the hole-doped cuprate systems studied so far with neutrons.

A second system to which a connection can be made is the electron-doped
superconductor Nd$_{2-x}$Ce$_x$CuO$_4$.  To see this, consider the measurements
of the instantaneous correlation length, or its inverse, $\kappa$, that have
been reported for Nd$_{1.85}$Ce$_{0.15}$CuO$_{4+\delta}$ \cite{mats92} and
La$_2$NiO$_{4.07}$ \cite{naka95}.  Such measurements involve an energy
integratation over the spin fluctuation cross section.  In the paramagnetic phase
of each of these systems,
$\kappa$ increases rapidly with temperature, differing substantially in each
case from behavior of the undoped parent compound.  The rapid variation of
$\kappa$ corresponds to the increase in $Q$ width found in our inelastic
measurements on the $\delta=0.105$ crystal. The change in the temperature
dependence of
$\kappa$ induced by doping has, in the former cases, been interpreted as
evidence for a weakening of the effective superexchange, $J$, based on
expectations for a 2D Heisenberg antiferromagnet.  In contrast, our
results for the $\delta=0.105$ sample suggest that an analysis based on a
spin-only model is likely to be inadequate, and one should consider the effect
of charge inhomogeneities.  The similarity in the temperature and
doping dependence of $\kappa$ in the electron-doped cuprate and the nickelate
suggests that one look for evidence of incipient charge stripes in as-grown
Nd$_{1.85}$Ce$_{0.15}$CuO$_{4+\delta}$ at $T\gtrsim200$~K.  Furthermore, the
abrupt appearance of superconductivity for a Ce concentration $x\gtrsim0.15$
might correspond to the stabilization of stripe correlations.  Experimental
tests are needed.

We gratefully acknowledge helpful discussions with V. J. Emery.  Work at 
Brookhaven was carried out under Contract
No.\ DE-AC02-76CH00016, Division of Materials Sciences, U.S. Department of
Energy.


\newlength{\zspace}
\settowidth{\zspace}{0}

\begin{table}
\caption{Summary of La$_2$NiO$_{4+\delta}$ samples, the nature of interstitial
and magnetic order (C$=$commensurate, IC$=$incommensurate), magnetic-ordering
temperature, and spin-wave velocity.}
\label{tb:g2e_par}
\begin{tabular}{dccdl}
$\delta$ & Interstitial & Magnetic & $T_m$\ \ \ & $\ \ \ \ \hbar c$ \\
         &    Order     &   Order   & (K)\ \ \ & (meV-\AA) \\[2pt]
\hline
\rule{0pt}{10pt}
0.00 & & C & 335. & 340\\
0.105 & 1D & C & 55. & \hspace{\zspace}$70\pm8$ \\
0.133 & 3D & IC & 110.5 & $200\pm20$ \\
\end{tabular}
\end{table}


\begin{figure}
\centerline{\psfig{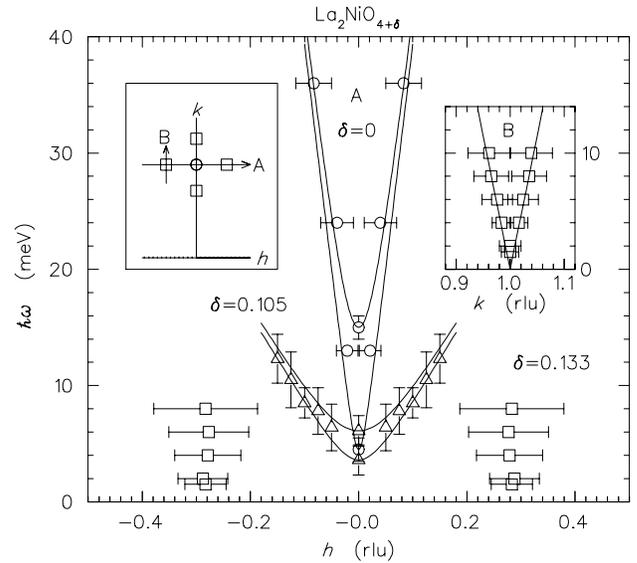}}
\medskip
\caption{Low-energy spin-wave dispersions measured on crystals of
La$_2$NiO$_{4+\delta}$ with $\delta=0$ (circles), 0.105 (triangles), and 0.133
(squares).  Left inset indicates directions $A$ and $B$ in the $(h,k,0)$
plane along which the dispersion has been characterized.  Main panel shows
dispersion along A; results of scans along $B$ for $\delta=0.133$ are shown in
the right inset.  Bars indicate measured peak widths (except at $h=0$ for
$\delta=0$), with no correction for resolution.
\label{fg:1}}
\end{figure}

\begin{figure}
\centerline{\psfig{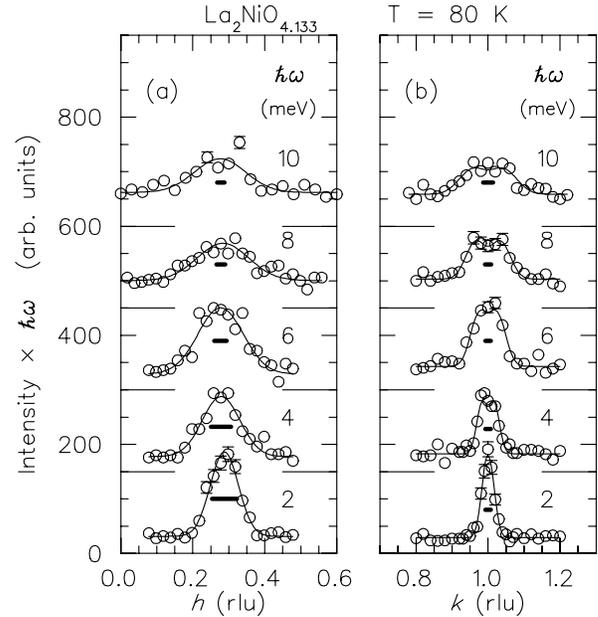}}
\medskip
\caption{Scans at constant energy transfer through the magnetic Bragg point
(0.278,1,0) of the $\delta=0.133$ crystal at $T=80$~K.  (a) Type-$A$ scans, as
defined in Fig.~1.  Lines through points are fits to a single Gaussian peak. 
(b) Type-$B$ scans.  Lines are fits of pairs of Gaussian peaks dispersing with
energy as indicated by the line in the right inset of Fig.~1.  Thick horizontal
bars indicate calculated resolution width.
\label{fg:2}}
\end{figure}

\begin{figure}
\centerline{\psfig{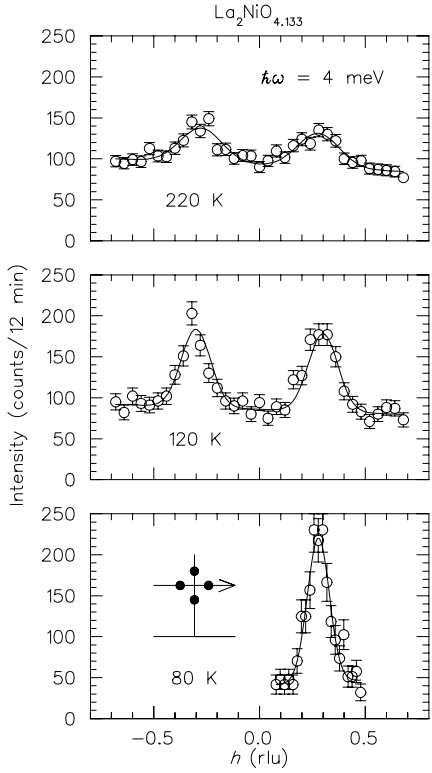}}
\medskip
\caption{Constant-energy scans at $\hbar\omega=4$~meV through the incommensurate
magnetic peak positions (as indicated in inset) at three temperatures: 80~K,
120~K and 220~K, the latter two being above the magnetic-order transition
(110.5~K).  The lines are fitted Gaussian peaks, symmetric about $h=0$.
\label{fg:3}}
\end{figure}

\begin{figure}
\centerline{\psfig{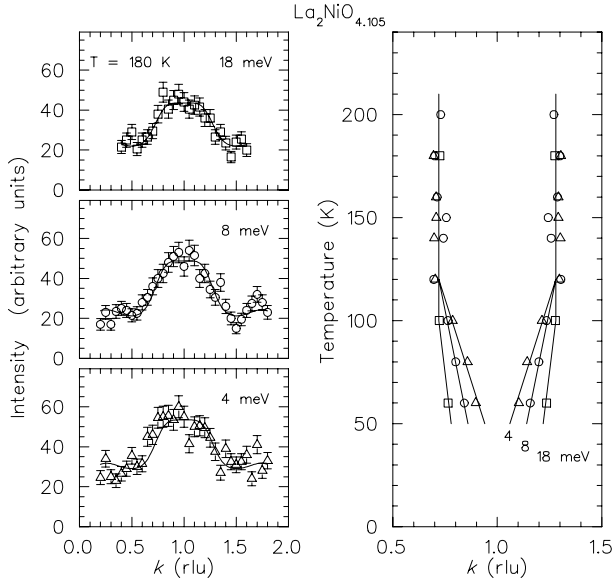}}
\medskip
\caption{Left: Constant-energy scans through the antiferromagnetic wavevector
measured on the $\delta=0.105$ crystal at $T=180$~K ($\approx3.6T_N$) with
$\hbar\omega=4$, 8, and 18~meV.  Lines are fits as described in the text. 
Right: Positions of half-maximum-intensity points (for peaks in constant-$E$
scans) vs.\ temperature for
$\hbar\omega=4$, 8, and 18~meV.  Lines are guides to the eye.
\label{fg:4}}
\end{figure}

\end{document}